\documentclass{article}

\usepackage{graphicx}

\begin{document}

\newcommand{\bfGamma}{\mbox{\boldmath $\it\Gamma$}}
\newcommand{\smallbfGamma}{\mbox{\boldmath $\scriptstyle \Gamma$}}
\newcommand{\bfq}{\mbox{\boldmath $\it q$}}
\newcommand{\bfp}{\mbox{\boldmath $\it p$}}
\newcommand{\bfn}{\mbox{\boldmath $\it n$}}



\begin{center}

{\LARGE \bf Lyapunov Spectra of Periodic Orbits  
for a Many-Particle System}

\end{center}

\vspace{0.3cm}
\begin{center}
{\it Tooru Taniguchi$^{1,2}$, Carl P. Dettmann$^{1}$ 
and Gary P. Morriss$^{2}$}

\vspace{0.2cm}
{$^{1}$ \it Department of Mathematics, University of Bristol, 
\\ University Walk, Bristol, BS8 1TW, United Kingdom}

\vspace{0.1cm}
{$^{2}$ \it School of Physics, University of New South Wales, 
Sydney, \\New South Wales 2052, Australia} 

\end{center}

\vspace{0.2cm}
\begin{quote}
   The Lyapunov spectrum corresponding to a periodic orbit 
for a two dimensional many particle system with hard core 
interactions is discussed. 
   Noting that the matrix to describe the tangent space dynamics 
has the block cyclic structure, 
the calculation of the Lyapunov spectrum is attributed to 
the eigenvalue problem of $16 \times 16$ reduced matrices 
regardless of the number of particles.  
   We show that there is the thermodynamic limit of the Lyapunov 
spectrum in this periodic orbit.  
   The Lyapunov spectrum has a step structure, which 
is explained by using symmetries of the reduced matrices.   
\end{quote}
\vspace{0.3cm}


\section{Introduction}

   Just as low dimensional chaos is revealed by a single positive 
Lyapunov exponent which indicates exponential sensitivity to initial 
conditions, systems with phase space of very high dimension can be 
characterized by their Lyapunov spectra which give information about 
many possible instabilities in the system.  
   As a concrete example, we consider the system consisting of $N$ 
disks with hard core interactions and periodic boundary conditions.
   This is a surprisingly good model of a fluid~\cite{EM}, and yet is
sufficiently simple that ergodic properties may be established under
fairly general conditions~\cite{S01}.

   The following is a brief review of the study of Lyapunov spectra 
in such systems; for more details and references in numerical work, see
Ref.~\cite{PH} and in analytical work, see Ref.~\cite{D00}.  
   It has been known for some time that the Lyapunov exponents of 
Hamiltonian systems come in plus/minus pairs, that is, the spectrum 
is symmetric about zero~\cite{AM}.  
   Non-equilibrium extensions were shown to exhibit symmetry about 
a point other than zero~\cite{ECM,DM96a}, leading to the discovery 
that these extensions also contain hidden Hamiltonian 
structure~\cite{DM96b,WL}. 
   Also known for more than twenty years is the algorithm for numerical 
computation of Lyapunov exponents due to Benettin and 
others~\cite{BGGS,SN}.  
   Later, a constraint method was introduced~\cite{HP}. 
   More recently, the effects of the hard collisions have been properly
taken into account~\cite{PH,del96}.  
   The existence of a thermodynamic limit in Lyapunov spectra, that is, 
that the spectrum retains its shape as the number of particles increases, 
has been put forward using random matrix approximations~\cite{N86},
numerical evidence~\cite{LPRV} and 
mathematical arguments~\cite{S96}, but recent numerical work has 
suggested in contrast, a logarithmic singularity of the largest Lyapunov 
exponent with the number of particles~\cite{SEI}.
   Lyapunov spectra for diatomic molecules (represented by dumb-bells)
show an explicit separation of the rotational and translational degrees
of freedom if the departure from sphericity is small enough~\cite{MPH}. 
   Finally, and of particular interest to us, careful simulations of
sufficiently large systems have revealed a step structure in the Lyapunov
spectrum for the {\em smallest} positive Lyapunov 
exponents~\cite{PH,del96,MPH}.
   The above references give an incomplete description in terms of ``Posch
Lyapunov modes'', phase space perturbations corresponding to these small
Lyapunov exponents which are approximately sinusoidal in position space.
   Recently Eckmann and Gat have suggested an explanation of the 
Lyapunov modes of a one dimensional system using a random matrix
approximation of the Lyapunov spectrum~\cite{EG}.  
   In this paper we study step structure in the Lyapunov spectrum 
of a two dimensional many particle system without making any approximations,
however we are restricted to periodic orbits.

   Periodic orbit theory~\cite{CAMTV,AAC} has proven very useful for
investigations of the corresponding low dimensional system, the periodic
Lorentz gas~\cite{CGS,V92,MR,CEG}, computing properties such as the
diffusion coefficient.  
   These methods cannot generally be applied directly to high 
dimensional systems due to the difficulty of finding all the periodic 
orbits, although a notable exception is the Kuramoto-Shivashinsky
PDE in a regime where the {\em effective} number of degrees of freedom
is small~\cite{CCP}.  
   However, periodic orbit arguments have been used to justify 
thermodynamic results such as non-negativity of the entropy
production~\cite{RM} and the Onsager relations~\cite{RC} without
explicitly finding any periodic orbits.  
   Periodic orbits of spatially extended systems in the form of 
coupled map lattices have been considered previously~\cite{AGGN,GA}, 
leading to block cyclic matrices similar to those observed in this paper.

   Here we apply methods similar to those to Ref.\cite{GA} to reduce 
the problem of computing the Lyapunov spectrum to that of finding the
eigenvalues of a relatively small ($16\times 16$) matrix.  
   A step structure is observed, which is related to the symmetries 
of the system.  
   Our formalism shows that, at least in the case of this periodic
orbit, the thermodynamic limit of the Lyapunov spectrum holds exactly.


\section{Lyapunov Exponents of Periodic Orbits for the  
Many-Particle System}

   The system which we consider in this paper is a two-dimensional 
Hamiltonian system consisting of $N$ disks,
interacting only by hard core collisions.
   We choose units such that the mass and radius of 
the particles are one. 
   We write the position and the momentum of the $j$-th particle
as $\bfq_{j}$ and $\bfp_{j}$, respectively, and for a later 
convenience we represent the phase space vector $\bfGamma$ as 
a column vector $(\bfq_{1}, 
\bfp_{1},\bfq_{2},\bfp_{2},\cdots,\bfq_{N},\bfp_{N})^{T}$ with 
$T$ the transpose operation.
     
   The dynamics of such a many-particle system is simply separated 
into the free flight part and the collision part, and the tangent 
vector $\delta\bfGamma_{n}$ of the phase space just after the $n$-th 
collision occurs is related to the tangent vector 
$\delta\bfGamma_{0}$ at the initial time by
$\delta\bfGamma_{n} = M_{n}\delta\bfGamma_{0}$ with 
the $4N\times 4N$ matrix $M_{n}$ represented as  

\begin{eqnarray}
   M_{n} \equiv M_{n}^{(c)}M_{n}^{(f)} 
   M_{n-1}^{(c)}M_{n-1}^{(f)} 
   \cdots M_{1}^{(c)}M_{1}^{(f)}.  
\label{Tange}\end{eqnarray}

\noindent Here $M_{j}^{(f)}$ is the $4N\times 4N$ matrix to 
specify the $j$-th  
free flight dynamics, and is given by 

\begin{eqnarray}
   M_{j}^{(f)} \equiv \mbox{Diag}(L_{1}^{(1)},L_{2}^{(1)}, 
   \cdots,L_{N}^{(1)})
\label{MatriMf}\end{eqnarray}

\noindent where $\mbox{Diag}(X_{1},X_{2},\cdots,X_{l})$ means 
the matrix on whose diagonal are the matrix blocks 
$X_{1},X_{2},\cdots,X_{l}$ for an integer $l$, and 
$L_{j}^{(1)}$ is defined by

\begin{eqnarray}
     L_{j}^{(1)} \equiv   
   \left(\begin{array}{cc}
      \underline{I} & \tau_{j}\underline{I}  \\
       \underline{0} & \underline{I} 
   \end{array}\right)
\label{MatriL1}\end{eqnarray}

\noindent where $\underline{I}$ and $\underline{0}$ are 
$2 \times 2$ identical and null matrices, respectively, 
and $\tau_{j}$ is its corresponding free flight time. 
   On the other hand, $M_{j}^{(c)}$ is the $4N\times 4N$ 
matrix to specify the $j$-th collision of particles.
   For the simplest case in which the $j$-th collision  involves only
the $k_j$-th and the $l_j$-th particles colliding,
   the matrix 
$M_{j}^{(c)}$ is given as the block matrix 

\begin{eqnarray}
   M_{j}^{(c)} \equiv   
   \left(\begin{array}{cccccccc}
      \tilde{I} &   &   &   &   &   &   & \\
        &   & \ddots &   &   &   &   &       \\
        &   &   & F_{j} &   & G_{j}  &   &       \\
        &   &   &   &  \ddots &   &    &       \\
        &   &   & G_{j}  &   &  F_{j} &    &         \\
        &   &   &   &   &   &   \ddots &     \\
        &   &   &   &   &   &   & \tilde{I}   
   \end{array}\right)
   \begin{array}{cccccccc}
       \\
       \leftarrow \;\;\;k_{j} \\
       \\
       \\
       \leftarrow \;\;\; l_{j}\\
       \\   
   \end{array}
\label{MatriMc}\end{eqnarray}
\vspace{-1.5cm} 
\begin{eqnarray}
   \begin{array}{cccccccc}
        &   &   \uparrow &&&   &  \uparrow    &  \\
        &   &   k_{j}  &&&   &  l_{j}    &              
   \end{array}
   \begin{array}{cccccccc}
       \\
       \\
       \\
       \\
       \\
       \\   
   \end{array} \nonumber
\label{MatriMcc}\end{eqnarray}
\vspace{-1.0cm}

\noindent where $\tilde{I}$ is the $4\times 4$ identity matrix and 
is put in the part of $\ddots$ of the above representation, 
and $F_{j}$ ($G_{j}$) is the $(k_j,k_j)$ and 
$(l_j,l_j)$ ($(k_j,l_j)$ and $(l_j,k_j)$) block matrix elements, 
and the $4\times 4$ null matrices are put in the other elements. 
   Here $4\times 4$ matrices $F_{j}$ and $G_{j}$ are defined by \cite{del96}   
    
\begin{eqnarray}
     F_{j} \equiv   
   \left(\begin{array}{cc}
      \underline{I}-L_{j}^{(2)} & \underline{0}  \\
      -L_{j}^{(3)} & \underline{I}-L_{j}^{(2)} 
   \end{array}\right)
\label{MatriF}\end{eqnarray}
\begin{eqnarray}
     G_{j} \equiv   
   \left(\begin{array}{cc}
      L_{j}^{(2)} & \underline{0}  \\
      L_{j}^{(3)} & L_{j}^{(2)} 
   \end{array}\right)
\label{MatriG}\end{eqnarray}   

\noindent with the $2\times 2$ matrices
$L_{j}^{(2)}$ and $L_{j}^{(3)}$ defined by

\begin{eqnarray}
   L_{j}^{(2)} \equiv \bfn_{j}\bfn_{j}^{T}
\label{MatriL2}\end{eqnarray}
\begin{eqnarray}
   L_{j}^{(3)} \equiv \bfn_{j}^{T}\Delta\bfp^{(j)} 
   \left( \underline{I} + \frac{\bfn_{j}\Delta\bfp^{(j)T}}
      {\bfn_{j}^{T}\Delta\bfp^{(j)}} \right)
   \left( \underline{I} - \frac{\Delta\bfp^{(j)}\bfn^{T}}
      {\Delta\bfp^{(j)T}\bfn_{j}}\right), 
\label{MatriL3}\end{eqnarray}   

\noindent (Note that all vectors in this paper are introduced as 
column vectors, so for example,  
$\bfn_{j}^{T}\Delta\bfp^{(j)}$ 
is a scalar and $\bfn_{j}\Delta\bfp^{(j)T}$ is a matrix.) 
where $\bfn_j$ is the unit 
vector pointing from the center of the $k_{j}$-th disk to the 
center of the $l_{j}$-th disk at the $j$-th collision, and 
$\Delta\bfp^{(j)}$ is the momentum difference $\bfp_{l_j}-\bfp_{k_j}$ 
just before the $j$-th collision.

   Now we consider the case that the movement of the system 
is periodic in time, so that the condition $\bfGamma_{n_{p}} 
=\bfGamma_{0}$ is satisfied for a integer $n_{p}$. 
   In this case the Lyapunov exponents $\lambda_{j}$, 
$j=1,2,\cdots,4N$ of the periodic orbit are defined by
$
   \lambda_{j} = (1/t(n_{p})) \ln |m_{j}(n_{p})|
$
\noindent in terms of the absolute value of the (generally complex)
eigenvalue $m_{j}(n_{p})$ of the matrix 
$M_{n_{p}}$, where $t(n_{p})\equiv\sum_{j=1}^{n_{p}}\tau_{j}$ 
is the period of this orbit.

   We put the set of the Lyapunov exponents in descending order, namely  
$\lambda_{1} \geq \lambda_{2} \geq \cdots \geq \lambda_{4N}$. 
   It is well known that in the time-independent 
Hamiltonian system the Lyapunov exponents satisfy the 
pairing rule~\cite{AM}, namely the condition $\lambda_{1}+\lambda_{4N}
=\lambda_{2}+\lambda_{4N-1}=\cdots=\lambda_{2N}+\lambda_{2N+1}=0$. 
   Noting this fact, thereafter we consider only the first half 
$\lambda_{1}$, $\lambda_{2}, \cdots, \lambda_{2N}$ 
of the full Lyapunov exponents, and refer their set as the Lyapunov 
spectrum in this paper.


\section{Periodic Orbit Model and the Reduced Matrix}

\begin{figure}[t]
\begin{center}
   \vspace{0cm}
   \includegraphics[width=10cm]{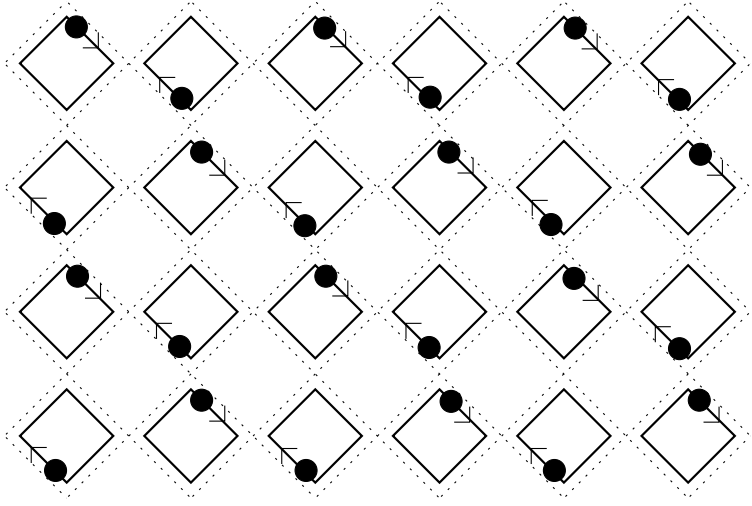}
   \vspace{0cm}
   \caption{Periodic orbit of the many particle system. 
      The circular dots show the positions of particles 
      at the initial time, and the solid lines give the
subsequent path in the direction of the arrows. 
\label{orbit}} 
   \vspace{0cm}
\end{center}
\end{figure}

   By using the method given in the previous section, 
we calculate the Lyapunov spectrum for the periodic orbit 
illustrated in Fig. \ref{orbit}. 
   In this periodic orbit each particle moves in a square 
orbit with the same absolute value of momentum and the same 
direction of rotation and with a constant free flight time 
$\tau$ (So the period of the orbit is $t(4)=4\tau$.). 
   We put $2N_{1}$ ($ 2N_{2}$) as the number of the particles
in each horizontal line (each vertical line) so that $N=4N_1N_2$.
We impose periodic boundary conditions, thus requiring the number
of particles in each direction to be even.

\begin{figure}[t]
\begin{center}
   \includegraphics[width=10cm]{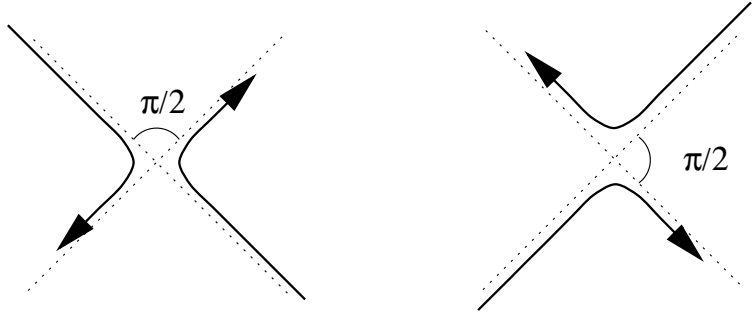}
   \vspace{0.5cm}
   \caption{Two types of particle collisions. 
      The solid lines give the path of the particles, which move
in the direction shown by the arrows.
\label{colli}} 
   \vspace{0cm}
\end{center}
\end{figure}

   In this periodic orbit model, there are only two types of 
collisions illustrated in Fig. \ref{colli}, in which the trajectories of 
two colliding particles are drawn with their moving directions 
shown by the arrows. 
   Now we consider the dynamics of 
particles for a free flight plus one of these two collisions.    
  Such a dynamics is described by the matrix multiplications
$\tilde{F}_{1} \equiv F_{1} \mbox{Diag}(L_{1}^{(1)},L_{1}^{(1)})$ and 
$\tilde{G}_{1}\equiv G_{1} \mbox{Diag}(L_{1}^{(1)},L_{1}^{(1)})$ 
($\tilde{F}_{2}\equiv F_{2} \mbox{Diag}(L_{2}^{(1)},L_{2}^{(1)})$ and 
$\tilde{G}_{2}\equiv G_{2} \mbox{Diag}(L_{2}^{(1)},L_{2}^{(1)})$) 
corresponding to the left (right) orbit in Fig. \ref{colli}. 
   By using Eqs. (\ref{MatriF}), (\ref{MatriG}), (\ref{MatriL2}) 
and (\ref{MatriL3}) these matrices are simply given by 

\begin{eqnarray}
   \tilde{F}_{1} = \left(\begin{array}{cccc}
       0 &  0 &   0    &   0    \\
       0 &  1 &   0    &  \tau  \\
      -v & -v & -v\tau & -v\tau \\
       v &  v &  v\tau & 1+v\tau  
   \end{array}\right)
\label{matriF1}\end{eqnarray}
\begin{eqnarray}
   \tilde{G}_{1} = \left(\begin{array}{cccc}
       1 &  0 &  \tau   &   0    \\
       0 &  0 &    0    &   0    \\
       v &  v & 1+v\tau & v\tau  \\
      -v & -v &  -v\tau & -v\tau  
   \end{array}\right)
\label{matriG1}\end{eqnarray}
\begin{eqnarray}
   \tilde{F}_{2} = \left(\begin{array}{cccc}
       1 &  0 &  \tau   &   0    \\
       0 &  0 &    0    &   0    \\
       v & -v & 1+v\tau & -v\tau  \\
       v & -v &  v\tau  & -v\tau  
   \end{array}\right)
\label{matriF2}\end{eqnarray}
\begin{eqnarray}
   \tilde{G}_{2} = \left(\begin{array}{cccc}
       0 &  0 &   0    &   0    \\
       0 &  1 &   0    &  \tau  \\
      -v &  v & -v\tau &  v\tau \\
      -v &  v & -v\tau & 1+v\tau  
   \end{array}\right), 
\label{matriG2}\end{eqnarray}

\noindent where one of the components of the collision 
vector is zero, namely $\bfn=(1,0)^{T}$ or $(0,1)^{T}$, 
and $v/\sqrt{2}$ is the speed of the particles.   

   The $4N\times 4N$ matrices $M_{j}^{(c)}M_{j}^{(f)}$, $j=1,2,3,4$ are 
represented as

\begin{eqnarray}
    M_{1}^{(c)}M_{1}^{(f)} 
   = \mbox{Diag}(P_{1},P_{2},P_{1},P_{2},\cdots,P_{1},P_{2})
\label{MatriM1}\end{eqnarray}
\begin{eqnarray}
    M_{2}^{(c)}M_{2}^{(f)} =  \left(\begin{array}{cccccccc}
      Q_{0}&Q_{1}&     &      &      &      &Q_{2} \\
      Q_{1}&Q_{0}&Q_{2}&      &      &      &      \\ 
           &Q_{2}&Q_{0}&Q_{1} &      &      &      \\
           &     &Q_{1}&Q_{0} &Q_{2} &      &      \\
           &     &     &\ddots&\ddots&\ddots&      \\
           &     &     &      &Q_{2} &Q_{0} &Q_{1} \\
      Q_{2}&     &     &      &      &Q_{1} &Q_{0}
   \end{array}\right)
\label{MatriM2}\end{eqnarray}
\begin{eqnarray}
   M_{3}^{(c)}M_{3}^{(f)} 
   = \mbox{Diag}(P_{2},P_{1},P_{2},P_{1},\cdots,P_{2},P_{1})
\label{MatriM3}\end{eqnarray}
\begin{eqnarray}
    M_{4}^{(c)}M_{4}^{(f)} = \left(\begin{array}{cccccccc}
      Q_{0}&Q_{2}&     &      &      &      &Q_{1} \\
      Q_{2}&Q_{0}&Q_{1}&      &      &      &      \\ 
           &Q_{1}&Q_{0}&Q_{2} &      &      &      \\
           &     &Q_{2}&Q_{0} &Q_{1} &      &      \\
           &     &     &\ddots&\ddots&\ddots&      \\
           &     &     &      &Q_{1} &Q_{0} &Q_{2} \\
      Q_{1}&     &     &      &      &Q_{2} &Q_{0}
   \end{array}\right)
\label{MatriM4}\end{eqnarray}

\noindent where the $8N_1\times8N_1$ matrices
$P_{j}$ and $Q_{j}$ are defined by  

\begin{eqnarray}
   P_{1} = \left(\begin{array}{cccccccccc}
      \tilde{F}_{1}&\tilde{G}_{1}&   &   &   &        &   &   & \\
      \tilde{G}_{1}&\tilde{F}_{0}&   &   &   &        &   &   & \\ 
           &     &\tilde{F}_{1}&\tilde{G}_{1} &   &   &   &   & \\
           &     &\tilde{G}_{1}&\tilde{F}_{1} &   &   &   &   & \\
           &     &     &      &\tilde{F}_{1}  &       &   &   & \\
           &     &     &      &      &\ddots&     &     & \\
           &     &     &      &      &      &\tilde{F}_{1}&     & \\
           &     &     &   &   &   &   &\tilde{F}_{1}&\tilde{G}_{1}\\
           &     &     &   &   &   &   &\tilde{G}_{1}&\tilde{F}_{1}
   \end{array}\right)
\label{MatriP1}\end{eqnarray}
\begin{eqnarray}
   P_{2} = \left(\begin{array}{cccccccccc}
      \tilde{F}_{1}&     &     &     &     &   &   &   &\tilde{G}_{1} \\
           &\tilde{F}_{0}&\tilde{G}_{1}&     &   &   &   &   & \\ 
           &\tilde{G}_{1}&\tilde{F}_{1}&     &     &   &   &   & \\
           &     &     &\tilde{F}_{1}&\tilde{G}_{1}&   &   &   & \\
           &     &     &\tilde{G}_{1}&\tilde{F}_{1}&   &   &   & \\
           &     &     &      &    &\ddots&     &     & \\
           &     &     &      &   &   &\tilde{F}_{1}&\tilde{G}_{1}& \\
           &     &     &      &   &   &\tilde{G}_{1}&\tilde{F}_{1}&\\
      \tilde{G}_{1}&     &     &   &    &    &    &    &\tilde{F}_{1}
   \end{array}\right)
\label{MatriP2}\end{eqnarray}

\begin{eqnarray}
   Q_{0} = 
   \mbox{Diag}(\tilde{F}_{2},\tilde{F}_{2},\tilde{F}_{2},\tilde{F}_{2},
   \cdots,\tilde{F}_{2},\tilde{F}_{2})
\label{MatriQ0}\end{eqnarray}
\begin{eqnarray}
   Q_{1} =
   \mbox{Diag}(\tilde{G}_{2},0,\tilde{G}_{2},0,\cdots,\tilde{G}_{2},0).
\label{MatriQ2}\end{eqnarray}
\begin{eqnarray}
   Q_{2} = 
   \mbox{Diag}(0,\tilde{G}_{2},0,\tilde{G}_{2},\cdots,0,\tilde{G}_{2})
\label{MatriQ1}\end{eqnarray}

\noindent In order to get these expression of the matrices 
the first horizontal row of particles is numbered
$1,2,\cdots,2N_{1}$ from left to
right, the second row numbered $2N_{1}+1,2N_{1}+2,\cdots,4N_{1}$
and so on until the last row numbered
$(2N_{2}-1)2N_{1}+1,(2N_{2}-1)2N_{1}+2,\cdots,4N_{2}N_{1}$. 
 
   The matrix $M_{n_{p}}$, whose eigenvalues lead to 
the Lyapunov spectrum,  is given by $M_{n_{p}}
=M_{4}^{(c)}M_{4}^{(f)} M_{3}^{(c)}M_{3}^{(f)} 
M_{2}^{(c)}M_{2}^{(f)}M_{1}^{(c)}M_{1}^{(f)}$.
 
   It is very important to note that this periodic model is invariant 
with respect to translations horizontally or vertically by the distance
corresponding to two particles. 
   This feature is reflected as block-cyclic structures in the matrix 
$M_{n_{p}}$. 
   By using this fact, as shown in Appendix A, we can show that 
the eigenvalues of the matrix $M_{n_{p}}$ are equal to the 
eigenvalues of the $16 \times 16$ matrices 
${\cal M}(2\pi n_{1}/N_{1},2\pi n_{2}/N_{2})$, 
$n_{1}=1,2,\cdots,N_{1}$, $n_{2}=1,2,\cdots,N_{2}$ 
with the matrix ${\cal M}(k,l)$ defined by  

\begin{eqnarray}
   \hspace{-1cm} && {\cal M}(k,l) \equiv \nonumber \\ 
   \hspace{-1cm} && \left(\begin{array}{cccc}
      S_{1}(-k,-l)       & S_{2}(-k,-l)         
             & T_{1}(k,l)e^{-il}  & T_{2}(k,l)e^{-i(k+l)} \\
      S_{2}(k,l)         & S_{1}(k,l)                         
             & T_{2}(-k,-l)e^{ik} & T_{1}(-k,-l)          \\ 
      T_{1}(-k,-l)e^{il} & T_{2}(-k,-l)e^{il}       
             & S_{1}(k,l)         & S_{2}(k,l)e^{-ik}     \\ 
      T_{2}(k,l)         & T_{1}(k,l)                                  
             & S_{2}(-k,-l)e^{ik} & S_{1}(-k,-l)
   \end{array}\right). 
\label{MatriCM}\end{eqnarray}

\noindent Here, $S_{1}(k,l)$, $S_{2}(k,l)$, $T_{1}(k,l)$ 
and $T_{2}(k,l)$ are defined by 

\begin{eqnarray}
   S_{1}(k,l) \equiv (\tilde{F}_{2}\tilde{F}_{1})^{2}
      +(\tilde{G}_{2}\tilde{G}_{1})^{2}
      +(\tilde{F}_{2}\tilde{G}_{1})^{2}e^{ik}
      +(\tilde{G}_{2}\tilde{F}_{1})^{2}e^{il}
\label{MatriS1}\end{eqnarray}
\begin{eqnarray}
   S_{2}(k,l) \equiv \tilde{F}_{2}\tilde{F}_{1}\tilde{F}_{2}\tilde{G}_{1}
      +\tilde{G}_{2}\tilde{G}_{1}\tilde{G}_{2}\tilde{F}_{1}
      +\tilde{F}_{2}\tilde{G}_{1}\tilde{F}_{2}\tilde{F}_{1}e^{ik}
      +\tilde{G}_{2}\tilde{F}_{1}\tilde{G}_{2}\tilde{G}_{1}e^{il}
\label{MatriS2}\end{eqnarray}
\begin{eqnarray}
   T_{1}(k,l) \equiv \tilde{F}_{2}\tilde{G}_{1}\tilde{G}_{2}\tilde{G}_{1}
      +\tilde{G}_{2}\tilde{F}_{1}\tilde{F}_{2}\tilde{F}_{1}
      +\tilde{G}_{2}\tilde{G}_{1}\tilde{F}_{2}\tilde{G}_{1}e^{ik}
      +\tilde{F}_{2}\tilde{F}_{1}\tilde{G}_{2}\tilde{F}_{1}e^{il}
\label{MatriT1}\end{eqnarray}
\begin{eqnarray}
   T_{2}(k,l) \equiv \tilde{F}_{2}\tilde{G}_{1}\tilde{G}_{2}\tilde{F}_{1}
      +\tilde{G}_{2}\tilde{F}_{1}\tilde{F}_{2}\tilde{G}_{1}
      +\tilde{G}_{2}\tilde{G}_{1}\tilde{F}_{2}\tilde{F}_{1}e^{ik}
      +\tilde{F}_{2}\tilde{F}_{1}\tilde{G}_{2}\tilde{G}_{1}e^{il}.
\label{MatriT2}\end{eqnarray} 

\noindent In the next section we investigate the Lyapunov spectrum 
of this periodic orbit model by using the eigenvalues of the matrix 
${\cal M}(k,l)$.

 
\section{Lyapunov Spectrum and its Step Structure}

   The reduced matrix (\ref{MatriCM}) is useful, not only to reduce 
the calculation time to get the full Lyapunov spectrum, but also 
to allow us to consider some properties of the Lyapunov spectrum 
itself. 
   One of important results obtained by such a consideration 
using the reduced matrix is the existence of the thermodynamic 
limit in the Lyapunov spectrum. 
   It should be noted that the $N_{1}N_{2}$ matrices
${\cal M}(k,l)$, with $k=2\pi n_1/N_1$, $l=2\pi n_2/N_2$ 
$n_{1}=1,2,\cdots,N_{1}$ and  $n_{2}=1,2,\cdots,N_{2}$ have 
the same form of the matrix given by Eq. (\ref{MatriCM}), so 
in the limit $N_{1}\rightarrow+\infty$ and $N_{2}\rightarrow +\infty$ 
the Lyapunov spectrum is given through the eigenvalues of the 
matrices ${\cal M}(k,l)$, $k\in [0, 2\pi)$ and $l\in [0, 2\pi)$. 
   This also implies that the maximum Lyapunov exponent takes a  
finite value in the thermodynamic limit in this periodic orbit model. 

\begin{figure}[t]
\begin{center}
   \vspace{0cm}
   \includegraphics[width=10.0cm]{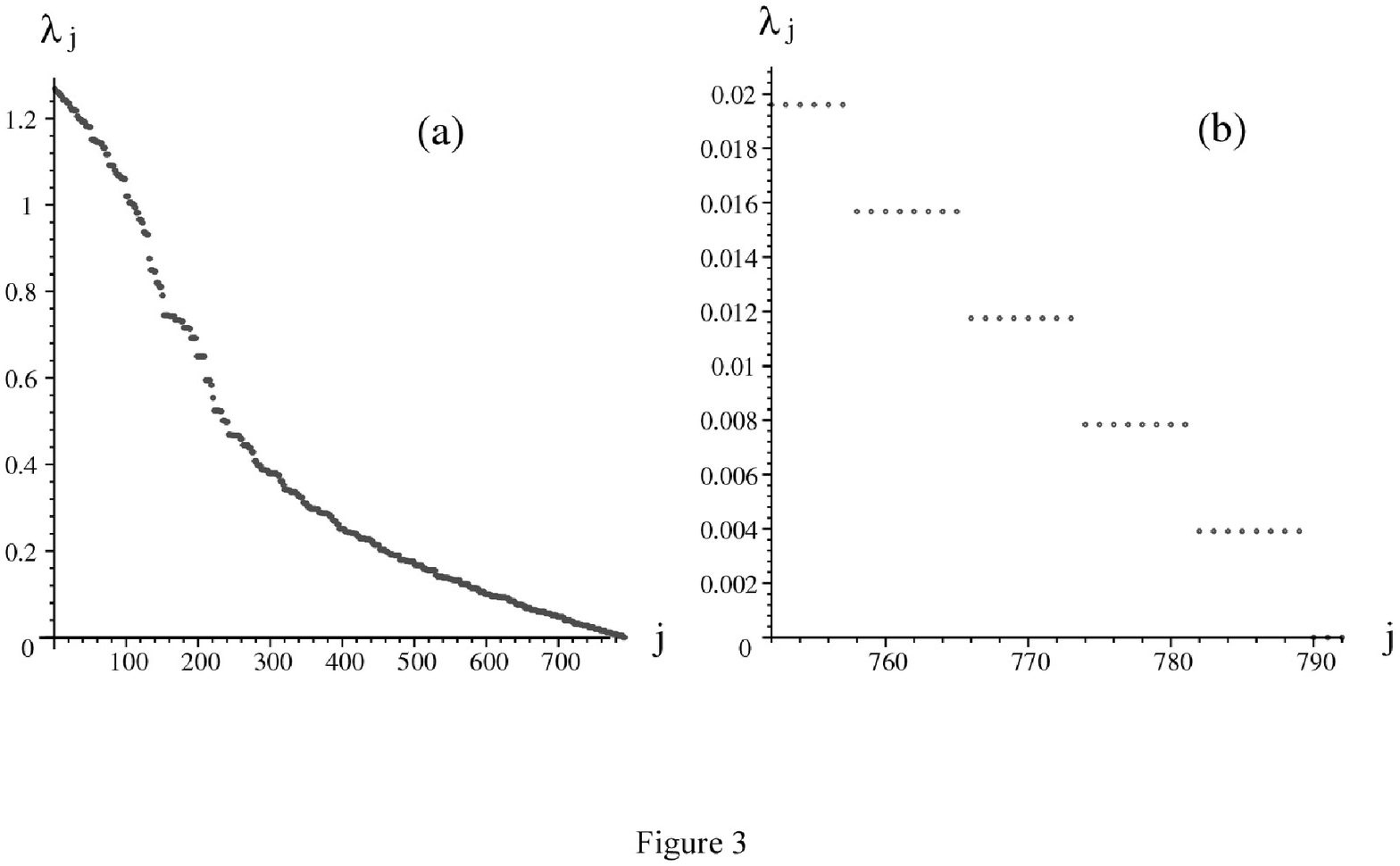}
   \vspace{0cm}
   \caption{Lyapunov spectrum 
   in the case of $(N_{1},N_{2})=(11,9)$. 
   (a) Full scale. (b) Large $j$ part.
	\label{lyaspe}} 
   \vspace{0cm}
\end{center}
\end{figure}

   Now we calculate the Lyapunov spectrum in our model. 
   Fig. \ref{lyaspe} is the Lyapunov spectrum in the case of 
$v=1.8$, $\tau=2.3$, $N_{1}=11$ and $N_{2}=9$ (So the 
total number of particles is $N=4N_1N_2=396$).
   One of the remarkable features of this Lyapunov spectrum is 
its step structure. 
   It is important to note that some steps of the Lyapunov spectrum 
are explained by using symmetries of the reduced matrix (\ref{MatriCM}). 
   Actually we can show (at least numerically) that in the case 
of Fig. \ref{lyaspe} the Lyapunov exponents calculated using the matrix 
${\cal M}(k,l)$ are invariant under the transformations $k\rightarrow
2\pi-k$ and $l\rightarrow2\pi-l$, leading to 
steps in the Lyapunov spectrum.  Performing both of these transformations
at once has the effect of taking the complex conjugate of $\cal M$ so
the result is obvious, however it is not immediately obvious that the
spectrum should be invariant if $k$ and $l$ are transformed separately.
 
\begin{figure}[t]
\begin{center}
   \vspace{0cm}
   \includegraphics[width=10.0cm]{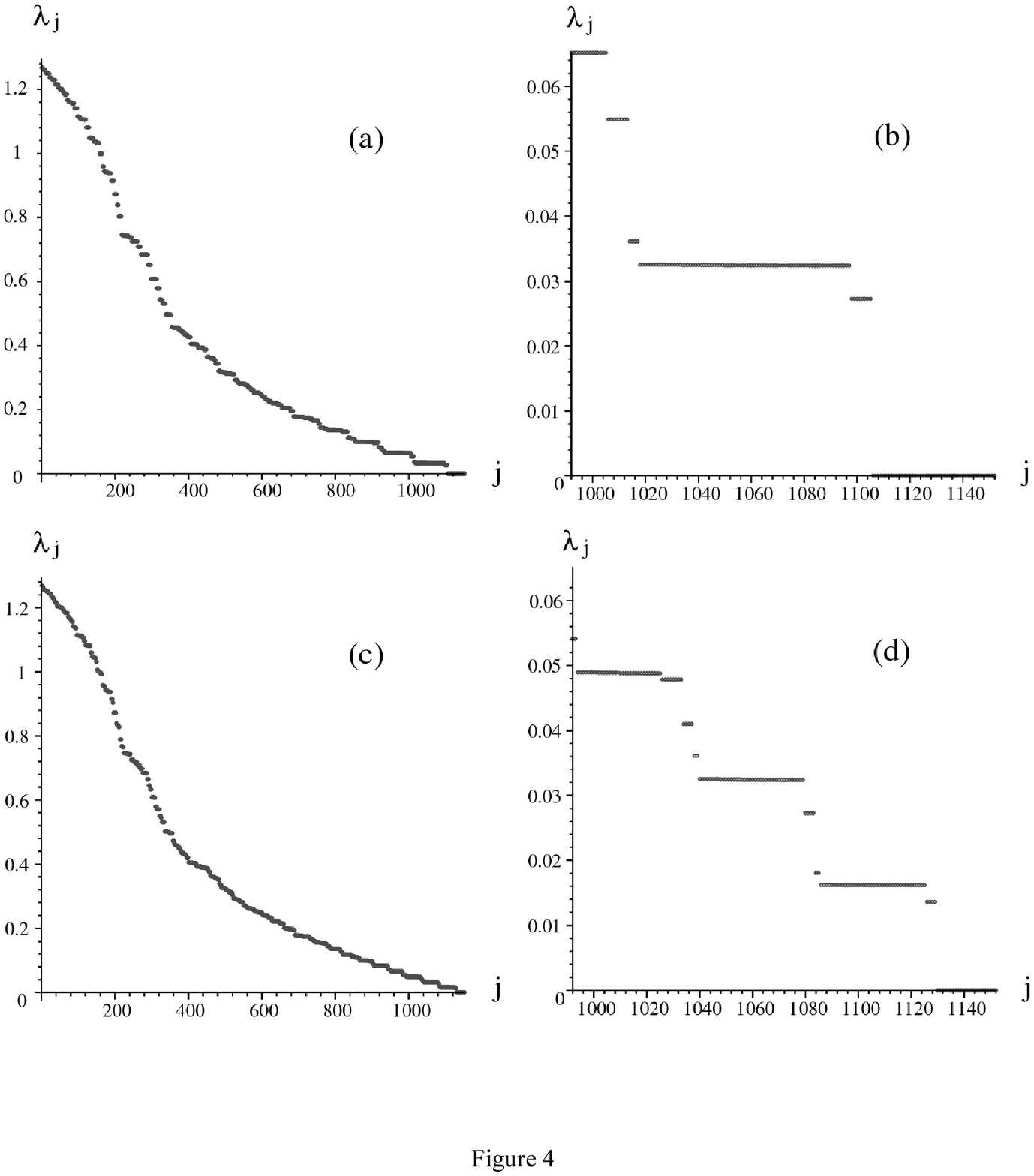}
   \vspace{0cm}
   \caption{Lyapunov spectra in the cases of a square system 
   $(N_{1},N_{2})=(12,12)$ ((a) Full scale. (b) Large $j$ part.) and 
   a rectangular system $(N_{1},N_{2})=(24,6)$ 
   ((c) Full scale. (d) Large $j$ part.).}
   \vspace{0cm}
\end{center}
\label{recsqu} 
\end{figure}

   We can investigate the relation between a symmetry of the system 
and the step structure in the Lyapunov spectrum another way. 
   For this purpose we consider the Lyapunov spectra in  
the square case and the rectangular case with the same  
number of particles.  
   Here, the square system has the symmetry 
for the exchange of the vertical and the horizontal directions, which 
the rectangular system does not have. 
   Fig. \ref{recsqu} is the Lyapunov spectra 
in the case of $(N_{1},N_{2})=(12,12)$ (the upper two graphs) 
and in the case $(N_{1},N_{2})=(24,6)$ (the lower two graphs) 
with $v=1.8$ and $\tau=2.3$. 
   These graphs show that there is not a remarkable difference 
in the global shapes of the Lyapunov spectrum in these two cases, 
but the square system has (even twice) longer 
steps in the Lyapunov spectrum than in the rectangular system,  
shown in Figs. \ref{recsqu} (b) and (d). 
   These longer steps come from an additional symmetry in 
the square system.
   Actually we can check numerically that the Lyapunov exponents
obtained from the reduced matrix (\ref{MatriCM}) are invariant under 
$k\leftrightarrow l$, which leads
to more degeneracy in the case $N_1=N_2=12$ than when $(N_1,N_2)=(24,6)$.
For $N_1=N_2$ all that is required is that $n_1$ and $n_2$ are interchanged.
For $(N_1,N_2)=(24,6)$ degeneracy by this mechanism only occurs if
$n_1$ is divisible by $4$ which is much rarer.

   It should be noted that some steps of the Lyapunov spectra 
are too close to be distinguished in Fig. \ref{recsqu}. 
   For example, if we can investigate more precisely, 
we can see 5 different steps in the long flat part of 
the Lyapunov exponents $\lambda_{j}$ in $j\in [1018,1097]$ 
(in $j\in [1086,1125]$) in the Lyapunov spectrum in 
the case of $(N_{1},N_{2})=(12,12)$ (in the case of 
$(N_{1},N_{2})=(24,6)$). 
   Similarly, most of the Lyapunov exponents $\lambda_{j}$, 
$j\in [1106,1152]$ ($j\in [1130,1152])$) in the case of 
$(N_{1},N_{2})=(12,12)$ (in the case of $(N_{1},N_{2})=(24,6)$) are 
not zero, just too small for the scale of the plot.

\section{Conclusion and Remarks}

   In this paper we have investigated the Lyapunov spectrum of a 
periodic orbit of a two-dimensional system, which consists 
of many disks with hard-core interactions. 
   The system has a rectangular shape (or a square shape 
in a special case), and we used periodic boundary conditions. 
   By the block cyclic structure of the matrix, 
which describes the tangent space dynamics, 
the calculation of the Lyapunov exponents is simplified to the 
eigenvalue problems of reduced $16 \times 16$ matrices however 
many particles. 
   The reduced matrix is used to consider the relation 
between such a step structure and symmetries of the system, 
and to show the existence of the thermodynamic limit in the 
Lyapunov spectrum.  
   In particular we showed that the difference of the aspect 
ratio of the system appears in the stepwise structure of the 
Lyapunov spectrum rather than in the global shape of the 
Lyapunov spectrum. 

   The Lyapunov spectrum of the periodic orbit discussed in this 
paper shows a step structure, but it should be noted that this step 
structure is different from the step structure obtained by 
the numerical work \cite{PH}. 
   More systematic investigations of the Lyapunov spectra of 
periodic orbits may be requested to explain the numerical results. 

   It should be emphasized that there are many periodic orbits 
in which we can calculate their Lyapunov spectra 
in many-hard-disk systems by using 
the block cyclic technique shown in this paper. 
   On the other hand we are still far from the 
position where we can calculate the general Lyapunov spectrum 
of the many hard disk system by using the periodic orbit 
expansion technique. 
   One of the problems is that in many particle system 
we could not know how to systematically find all of 
the periodic orbits whose periods are less than a given length. 
   In addition, the periodic orbits of many-panicle system 
are typically distributed continuously, not isolated 
from each other.  
   These problems make the application of the periodic orbit expansion 
technique to the many-particle systems difficult, and not yet solved. 

\section*{Acknowledgements}
GM and TT are grateful for financial support from the
Australian Research Council. CD is grateful for financial
support from the Nuffield Foundation, grant NAL/00353/G.


\appendix

\section{Block Cyclic Structure and the Reduced Matrix}

   In this appendix we show that the eigenvalues of the 
matrix $M_{n_{p}}$ are equal to the eigenvalues of the 
matrices ${\cal M}(2\pi n_{1}/N_{1},2\pi n_{2}/N_{2})$, 
$n_{1}=1,2,\cdots,N_{1}$, $n_{2}=1,2,\cdots,N_{2}$, 
given by Eq. (\ref{MatriCM}). 

   First we calculate the matrix $M_{n_{p}}$. 
   We multiply the matrices 
$M_{j}^{(c)}M_{j}^{(f)}$, $j=1,2,3,4$ given by Eqs. 
(\ref{MatriM1}), (\ref{MatriM2}), (\ref{MatriM3}) and 
(\ref{MatriM4}), and obtain 

\begin{eqnarray}
   M_{n_{p}} = \left(\begin{array}{cccccc}
      K_{0}&K_{1}&     &      &      &K_{2} \\
      K_{2}&K_{0}&K_{1}&      &      &      \\ 
           &K_{2}&K_{0}&K_{1} &      &      \\
           &     &\ddots&\ddots&\ddots&      \\
           &     &      &K_{2} &K_{0} &K_{1} \\
      K_{1}&     &      &      &K_{2} &K_{0}
   \end{array}\right)
\label{MatriMnp}\end{eqnarray}

\noindent where $K_{j}$, $j=0,1,2$ are defined by 

\begin{eqnarray}
    K_{0} \equiv \left(\begin{array}{cc}
      L_{02}L_{01}+L_{21}L_{11}+L_{11}L_{21} & 
         L_{02}L_{12}+L_{21}L_{02}  \\
      L_{01}L_{11}+L_{22}L_{01} & 
         L_{01}L_{02}+L_{22}L_{12}+L_{12}L_{22}
   \end{array}\right)
\label{MatriK0}\end{eqnarray}
\begin{eqnarray}
    K_{1} \equiv \left(\begin{array}{cc}
      L_{21}^{2} &  \tilde{0}   \\
      L_{01}L_{21}+L_{12}L_{01} & L_{12}^{2}
   \end{array}\right)
\label{MatriK1}\end{eqnarray}
\begin{eqnarray}
    K_{2} \equiv \left(\begin{array}{cc}
      L_{11}^{2} & L_{11}L_{02}+L_{02}L_{22} \\
      \tilde{0} & L_{22}^{2}
   \end{array}\right)
\label{MatriK2}\end{eqnarray}

\noindent with the matrices $L_{jk}\equiv Q_{j}P_{k}$ and $8N_{1} \times 
8N_{1}$ null matrix $\tilde{0}$. 
   Eq. (\ref{MatriMnp}) implies that the matrix $M_{n_{p}}$ 
has the block cyclic structure, so it is block-diagonalized 
by the orthogonal matrix $U(16N_{1},N_{2})$ introduced through

\begin{eqnarray}
    U(j,k) \equiv \frac{1}{\sqrt{k}} \left(\begin{array}{ccccc}
      I_{j}e^{2\pi i 1\times 1/k} &  I_{j}e^{2\pi i 1\times 2/k} & 
           \cdots &  I_{j}e^{2\pi i 1\times k/k} \\
      I_{j}e^{2\pi i 2\times 1/k} &  I_{j}e^{2\pi i 2\times 2/k} & 
           \cdots &  I_{j}e^{2\pi i 2\times k/k} \\
      \vdots & \vdots & & \vdots \\
      I_{j}e^{2\pi i k\times 1/k} &  I_{j}e^{2\pi i k\times 2/k} & 
           \cdots &  I_{j}e^{2\pi i k\times k/k} 
   \end{array}\right)
\label{MatriU}\end{eqnarray}

\noindent with the $j \times j$ identical matrix $I_{j}$ so that 
we obtain the matrix $U(16N_{1},N_{2})^{\dagger}$ $M_{n_{p}}U(16N_{1},N_{2})$ 
$=\mbox{Diag}(A(2\pi1/N_{2}),$ $
A(2\pi2/N_{2}),\cdots, A(2\pi N_{2}/N_{2})$ with ${\dagger}$ meaning 
to take its Hermitian conjugate. 
   Here the matrix $A(l)$ is defined by  

\begin{eqnarray}
    A(l) \equiv K_{0} + K_{1}e^{il} + K_{2}e^{-il}.  
\label{MatriA}\end{eqnarray}

\noindent The eigenvalues of the 
matrix $M_{n_{p}}$ are equal to the eigenvalues of the 
matrices $A(2\pi n_{2}/N_{2})$, $n_{2}=1,2,\cdots,N_{2}$.

   The matrix $A(l)$ is represented as  

\begin{eqnarray}
    A(l) = \left(\begin{array}{cc}
      B^{(1)}(l) & B^{(2)}(l) \\ B^{(3)}(l) & B^{(4)}(l)
    \end{array}\right) 
\label{MatriA2}\end{eqnarray}

\noindent where $B^{(j)}(l)$ is defined by 

\begin{eqnarray}
   B^{(j)}(l) \equiv \left(\begin{array}{cccccc}
      B^{(j)}_{0}(l)&B^{(j)}_{1}(l)&     &      &      &B^{(j)}_{2}(l) \\
      B^{(j)}_{2}(l)&B^{(j)}_{0}(l)&B^{(j)}_{1}(l)&      &      &      \\ 
           &B^{(j)}_{2}(l)&B^{(j)}_{0}(l)&B^{(j)}_{1}(l)&      &      \\
           &     &\ddots&\ddots&\ddots&      \\
           &     &      &B^{(j)}_{2}(l)&B^{(j)}_{0}(l)&B^{(j)}_{1}(l) \\
      B^{(j)}_{1}(l)&     &      &      &B^{(j)}_{2}(l) &B^{(j)}_{0}(l)
   \end{array}\right).
\label{MatriB}\end{eqnarray}

\noindent Here the matrices $B^{(j)}_{k}(l)$, $j=1,2,3,4$ and 
$k=0,1,2$ are defined by 

\begin{eqnarray}
   B^{(1)}_{0}(l) &\equiv& \left(\begin{array}{c}
      (\tilde{F}_{2}\tilde{F}_{1})^{2}
         +(\tilde{G}_{2}\tilde{G}_{1})^{2} 
         +(\tilde{G}_{2}\tilde{F}_{1})^{2}e^{-il} \\ 
      \tilde{F}_{2}\tilde{F}_{1}\tilde{F}_{2}\tilde{G}_{1}
         +\tilde{G}_{2}\tilde{G}_{1}\tilde{G}_{2}\tilde{F}_{1}
         +\tilde{G}_{2}\tilde{F}_{1}\tilde{G}_{2}\tilde{G}_{1}e^{il} 
   \end{array}\right. \nonumber \\  
   &&\hspace{1cm} \left.\begin{array}{c}
      \tilde{F}_{2}\tilde{F}_{1}\tilde{F}_{2}\tilde{G}_{1}
         +\tilde{G}_{2}\tilde{G}_{1}\tilde{G}_{2}\tilde{F}_{1}
         +\tilde{G}_{2}\tilde{F}_{1}\tilde{G}_{2}\tilde{G}_{1}e^{-il} \\ 
      (\tilde{F}_{2}\tilde{F}_{1})^{2}+(\tilde{G}_{2}\tilde{G}_{1})^{2} 
         +(\tilde{G}_{2}\tilde{F}_{1})^{2}e^{il} 
   \end{array}\right) 
\label{MatriB10}\end{eqnarray}
\begin{eqnarray}
   B^{(1)}_{1}(l) \equiv \left(\begin{array}{cc}
      \hat{0} & \hat{0} \\
      \tilde{F}_{2}\tilde{G}_{1}\tilde{F}_{2}\tilde{F}_{1} 
      & (\tilde{F}_{2}\tilde{G}_{1})^{2} \\ 
   \end{array}\right)
\label{MatriB11}\end{eqnarray}
\begin{eqnarray}
   B^{(1)}_{2}(l) \equiv \left(\begin{array}{cc}
      (\tilde{F}_{2}\tilde{G}_{1})^{2} & 
      \tilde{F}_{2}\tilde{G}_{1}\tilde{F}_{2}\tilde{F}_{1}\\
      \hat{0} & \hat{0} \\ 
   \end{array}\right)
\label{MatriB12}\end{eqnarray}

\begin{eqnarray}
   B^{(2)}_{0}(l) &\equiv& \left(\begin{array}{c}
      \tilde{F}_{2}\tilde{F}_{1}\tilde{G}_{2}\tilde{F}_{1} 
      +(\tilde{G}_{2}\tilde{F}_{1}\tilde{F}_{2}\tilde{F}_{1}  
         +\tilde{F}_{2}\tilde{G}_{1}\tilde{G}_{2}\tilde{G}_{1}) e^{-il} \\ 
      \tilde{G}_{2}\tilde{G}_{1}\tilde{F}_{2}\tilde{F}_{1} 
   \end{array}\right. \nonumber \\  
   &&\hspace{1cm} \left.\begin{array}{c}
      \tilde{G}_{2}\tilde{G}_{1}\tilde{F}_{2}\tilde{F}_{1} e^{-il}\\ 
      \tilde{F}_{2}\tilde{G}_{1}\tilde{G}_{2}\tilde{G}_{1}
         +\tilde{G}_{2}\tilde{F}_{1}\tilde{F}_{2}\tilde{F}_{1}  
         +\tilde{F}_{2}\tilde{F}_{1}\tilde{G}_{2}\tilde{F}_{1} e^{-il} 
   \end{array}\right) 
\label{MatriB20}\end{eqnarray}
\begin{eqnarray}
   B^{(2)}_{1}(l) \equiv \left(\begin{array}{cc}
      \tilde{G}_{2}\tilde{G}_{1}\tilde{F}_{2}\tilde{G}_{1} e^{-il} & 
      \hat{0} \\
      \tilde{F}_{2}\tilde{G}_{1}\tilde{G}_{2}\tilde{F}_{1} 
         +\tilde{G}_{2}\tilde{F}_{1}\tilde{F}_{2}\tilde{G}_{1} 
         + \tilde{F}_{2}\tilde{F}_{1}\tilde{G}_{2}\tilde{G}_{1} e^{-il} & 
      \hat{0} \\ 
   \end{array}\right)
\label{MatriB21}\end{eqnarray}
\begin{eqnarray}
   B^{(2)}_{2}(l) \equiv \left(\begin{array}{cc}
      \hat{0} & \tilde{F}_{2}\tilde{F}_{1}\tilde{G}_{2}\tilde{G}_{1} 
         +(\tilde{G}_{2}\tilde{F}_{1}\tilde{F}_{2}\tilde{G}_{1} 
         + \tilde{F}_{2}\tilde{G}_{1}\tilde{G}_{2}\tilde{F}_{1}) e^{-il} \\ 
      \hat{0} &  \tilde{G}_{2}\tilde{G}_{1}\tilde{F}_{2}\tilde{G}_{1} \\ 
   \end{array}\right)
\label{MatriB22}\end{eqnarray}

\begin{eqnarray}
   B^{(3)}_{0}(l) &\equiv& \left(\begin{array}{c}
      \tilde{F}_{2}\tilde{F}_{1}\tilde{G}_{2}\tilde{F}_{1}
         +(\tilde{F}_{2}\tilde{G}_{1}\tilde{G}_{2}\tilde{G}_{1} 
         +\tilde{G}_{2}\tilde{F}_{1}\tilde{F}_{2}\tilde{F}_{1}) e^{il} \\ 
      \tilde{F}_{2}\tilde{G}_{1}\tilde{G}_{2}\tilde{F}_{1}
         +\tilde{G}_{2}\tilde{F}_{1}\tilde{F}_{2}\tilde{G}_{1}
         +\tilde{F}_{2}\tilde{F}_{1}\tilde{G}_{2}\tilde{G}_{1}e^{il} 
   \end{array}\right. \nonumber \\  
   &&\hspace{1cm} \left.\begin{array}{c}
      \tilde{F}_{2}\tilde{F}_{1}\tilde{G}_{2}\tilde{G}_{1}
         +(\tilde{F}_{2}\tilde{G}_{1}\tilde{G}_{2}\tilde{F}_{1}
         +\tilde{G}_{2}\tilde{F}_{1}\tilde{F}_{2}\tilde{G}_{1})e^{il} \\ 
      \tilde{F}_{2}\tilde{G}_{1}\tilde{G}_{2}\tilde{G}_{1}
         +\tilde{G}_{2}\tilde{F}_{1}\tilde{F}_{2}\tilde{F}_{1} 
         +\tilde{F}_{2}\tilde{F}_{1}\tilde{G}_{2}\tilde{F}_{1}e^{il} 
   \end{array}\right) 
\label{MatriB30}\end{eqnarray}
\begin{eqnarray}
   B^{(3)}_{1}(l) \equiv \left(\begin{array}{cc}
      \hat{0} & \hat{0} \\
      \tilde{G}_{2}\tilde{G}_{1}\tilde{F}_{2}\tilde{F}_{1} & 
      \tilde{G}_{2}\tilde{G}_{1}\tilde{F}_{2}\tilde{G}_{1} \\ 
   \end{array}\right)
\label{MatriB31}\end{eqnarray}
\begin{eqnarray}
   B^{(3)}_{2}(l) \equiv \left(\begin{array}{cc}
      \tilde{G}_{2}\tilde{G}_{1}\tilde{F}_{2}\tilde{G}_{1}e^{il} & 
      \tilde{G}_{2}\tilde{G}_{1}\tilde{F}_{2}\tilde{F}_{1} e^{il}\\
      \hat{0} & \hat{0} \\ 
   \end{array}\right)
\label{MatriB32}\end{eqnarray}

\begin{eqnarray}
   B^{(4)}_{0}(l) &\equiv& \left(\begin{array}{c}
      (\tilde{F}_{2}\tilde{F}_{1})^{2}+(\tilde{G}_{2}\tilde{G}_{1})^{2} 
         +(\tilde{G}_{2}\tilde{F}_{1})^{2}e^{il} \\ 
      \tilde{F}_{2}\tilde{G}_{1}\tilde{F}_{2}\tilde{F}_{1} 
   \end{array}\right. \nonumber \\  
   &&\hspace{1cm} \left.\begin{array}{c}
      \tilde{F}_{2}\tilde{G}_{1}\tilde{F}_{2}\tilde{F}_{1} \\ 
      (\tilde{F}_{2}\tilde{F}_{1})^{2}+(\tilde{G}_{2}\tilde{G}_{1})^{2} 
         +(\tilde{G}_{2}\tilde{F}_{1})^{2}e^{-il} 
   \end{array}\right) 
\label{MatriB40}\end{eqnarray}
\begin{eqnarray}
   B^{(4)}_{1}(l) \equiv \left(\begin{array}{cc}
      (\tilde{F}_{2}\tilde{G}_{1})^{2} & \hat{0} \\
      \tilde{F}_{2}\tilde{F}_{1}\tilde{F}_{2}\tilde{G}_{1}
         +\tilde{G}_{2}\tilde{G}_{1}\tilde{G}_{2}\tilde{F}_{1} 
         + \tilde{G}_{2}\tilde{F}_{1}\tilde{G}_{2}\tilde{G}_{1} e^{-il} & 
      \hat{0} \\ 
   \end{array}\right)
\label{MatriB41}\end{eqnarray}
\begin{eqnarray}
   B^{(4)}_{2}(l) \equiv \left(\begin{array}{cc}
      \hat{0} & \tilde{F}_{2}\tilde{F}_{1}\tilde{F}_{2}\tilde{G}_{1} 
         +\tilde{G}_{2}\tilde{G}_{1}\tilde{G}_{2}\tilde{F}_{1} 
         + \tilde{G}_{2}\tilde{F}_{1}\tilde{G}_{2}\tilde{G}_{1} e^{il} \\ 
      \hat{0} &  (\tilde{F}_{2}\tilde{G}_{1})^{2}  
   \end{array}\right)
\label{MatriB42}\end{eqnarray}

\noindent with the $4 \times 4$ null matrix $\hat{0}$. 
   It follows from Eqs. (\ref{MatriU}), (\ref{MatriA2}) and 
(\ref{MatriB}) that  

\begin{eqnarray}
   \hspace{-2cm} && \mbox{Diag}(U(8,N_{1}),U(8,N_{1}))^{\dagger} 
      A(l) \; \mbox{Diag}(U(8,N_{1}),U(8,N_{1})) \nonumber \\ 
   && =  
   \left(\begin{array}{cc}
      C^{(1)}(l) & C^{(2)}(l) \\ C^{(3)}(l) & C^{(4)}(l)
   \end{array}\right)
\label{MatriA3}\end{eqnarray}

\noindent where $C^{(j)}(l)$, $j=1,2,3,4$ are defined by 
$C^{(j)}(l) 
\equiv \mbox{Diag}($ $D^{(j)}(2\pi 1/N_{1},l),$ 
$D^{(j)}(2\pi 2/N_{1},l), $ $
\cdots,D^{(j)}(2\pi N_{1}/N_{1},l) $ $)$ with $D^{(j)}(k,l) 
\equiv B^{(j)}_{0}$ $+B^{(j)}_{1}e^{ik}$ $+B^{(j)}_{2}e^{-ik}$. 
   The matrix ${\cal M}(k,l)$ is introduced as 

\begin{eqnarray}
   {\cal M}(k,l) =   
   \left(\begin{array}{cc}
      D^{(1)}(k,l) & D^{(2)}(k,l) \\ D^{(3)}(k,l) & D^{(4)}(k,l)
   \end{array}\right), 
\label{MatriCM2}\end{eqnarray}
 
\noindent which is equal to Eq. (\ref{MatriCM}). 
   Therefore the eigenvalues of the 
matrix $A(l)$ are equal to the eigenvalues of the 
matrices ${\cal M}(2\pi n_{1}/N_{1},l)$, 
$n_{1}=1,2,\cdots,N_{1}$. 
   It implies that the eigenvalues of the 
matrix $M_{n_{p}}$ are equal to the eigenvalues of the 
matrices ${\cal M}(2\pi n_{1}/N_{1},$ $2\pi n_{2}/N_{2})$, 
$n_{1}=1,2,\cdots,N_{1}$, $n_{2}=1,2,\cdots,N_{2}$.



\end{document}